\documentstyle[preprint,prb,aps]{revtex}
\begin{document}

\begin{titlepage}

\title
{Magnetization quantum tunneling at excited levels for biaxial 
spin system in an arbitrarily
directed magnetic field}

\author{Rong L\"{u}\footnote {Author to whom 
the correspondence should be addressed.\\
Electronic address: rlu@castu.phys.tsinghua.edu.cn}, Su-Peng Kou,
Jia-Lin Zhu, Lee Chang, and Bing-Lin Gu} 
\address{Center for Advanced Study, Tsinghua University,
Beijing 100084, P. R. China}

\maketitle
\begin{abstract}
The quantum
tunneling of the magnetization vector between excited levels
are studied theoretically in
single-domain ferromagnetic nanoparticles with biaxial crystal symmetry
placed in an external magnetic field at an arbitrarily directed angle in
the  $ZX$ plane.
By applying the periodic instanton method in the
spin-coherent-state path-integral representation, we
calculate the tunnel splittings and the
tunneling rates between excited levels in the low barrier
limit for different angle ranges of the external magnetic field
($\theta_{H}=\pi/2$, $\pi/2\ll\theta_{H}\ll\pi$, and $\theta_{H}=\pi$).
The temperature dependences of the tunneling frequency and
the decay rate are clearly shown for each case.
Our results show that the tunnel splittings and the
tunneling rates depend on the 
orientation of the external magnetic
field distinctly, which provides a possible experimental
test for magnetic quantum tunneling in
nanometer-scale single-domain ferromagnets.

\noindent
{\bf PACS number(s)}:  75.45.+j, 75.50.Ee
\end{abstract}

\end{titlepage}

\section*{I. Introduction}

Macroscopic quantum tunneling (MQT) and coherence (MQC) of the magnetization
were intensively investigated both theoretically and experimentally in
recent years.\cite{1} More recently, much attention was attracted to the
spin tunneling in the single-domain ferromagnetic (FM) nanoparticles in the
presence of an external magnetic field applied at an arbitrary angle. The
MQT problem for FM particles with uniaxial crystal symmetry was first
studied by Zaslavskii with the help of mapping the spin system onto a
one-dimensional particle system.\cite{2} For the same crystal symmetry,
Miguel and Chudnovsky\cite{3} calculated the tunneling rate by applying the
imaginary-time path integral, and demonstrated that the angular and field
dependences of the tunneling exponent obtained by Zaslavskii's method and by
the path-integral method coincide precisely. Kim and Hwang performed a
calculation based on the instanton technique for FM particles with biaxial
and tetragonal crystal symmetry.\cite{4} Kim extended the tunneling rate for
biaxial crystal symmetry to a finite temperature, and presented the
numerical results for the WKB exponent below the crossover temperature and
their approximate formulas around the crossover temperature.\cite{5} The
quantum-classical transition of the escape rate for FM particles with
uniaxial crystal symmetry in an arbitrarily directed field was studied by
Garanin, Hidalgo and Chudnovsky with the help of mapping onto a particle
moving in a double-well potential.\cite{6} The switching field measurement
was carried out on single-domain FM nanoparticles of Barium ferrite
(BaFeCoTiO) containing about $10^5-10^6$ spins.\cite{7} The measured angular
dependance of the crossover temperature was found to be in excellent
agreement with the theoretical prediction,\cite{3} which strongly suggests
the MQT of magnetization in the BaFeCoTiO nanoparticles. L\"{u} {\it et al}.
studied the MQT and MQC of the N\'{e}el vector in single-domain
antiferromagnetic (AFM) nanoparticles with biaxial, tetragonal, and
hexagonal crystal symmetry in an arbitrarily directed field.\cite{8}

It is noted that the previous results of MQT of the magnetization vector at
excited levels in an arbitrarily directed field were obtained by numerically
solving the equation of motion satisfied by the least trajectory.\cite{5}
The purpose of this paper is to present an analytical investigation of the
quantum tunneling at excited levels in the biaxial FM particles in an
arbitrarily directed field, based on the periodic instanton method.\cite
{9,10} Both the nonvacuum (or thermal) instanton and bounce solution, the
WKB exponents and the preexponential factors are evaluated exactly for
different angle ranges of the magnetic field $(\theta _H=\pi /2$, $\pi
/2+O\left( \epsilon ^{3/2}\right) \ <\theta _H<\pi -O(\epsilon ^{3/2})$, and 
$\theta _H=\pi )$. Our results show that the distinct angular dependence,
together with the dependence of the WKB tunneling rate on the strength of
the external magnetic field, may provide an independent experimental test
for the magnetic tunneling at excited levels in single-domain FM
nanoparticles.

This paper is structured in the following way. In Sec. II, we review briefly
some basic ideas of MQT and MQC in FM\ particles. And we discuss the
fundamentals concerning the computation of level splittings and tunneling
rates of excited states in the double-well-like potential. In Secs. III, we
study the spin tunneling at excited levels for FM particles with biaxial
crystal symmetry in the presence of an external magnetic field applied in
the $ZX$ plane with a range of angles $\pi /2\leq \theta _H\leq \pi $. The
conclusions are presented in Sec. V.

\section*{II. MQT and MQC of the magnetization vector in FM particles}

The system of interest is a nanometer-scale single-domain ferromagnet at a
temperature well below its anisotropy gap. For such a FM particle, the
tunnel splitting for MQC or the tunneling rate for MQT is determined by the
imaginary-time transition amplitude from an initial state $\left|
i\right\rangle $ to a final state $\left| f\right\rangle $ as 
\begin{equation}
U_{fi}=\left\langle f\right| e^{-HT}\left| i\right\rangle =\int D\Omega \exp
\left( -S_E\right) ,  \eqnum{1}
\end{equation}
where $S_E$ is the Euclidean action and $D\Omega $ is the measurement of the
path integral. In the spin-coherent-state representation, the Euclidean
action can be expressed as 
\begin{equation}
S_E\left( \theta ,\phi \right) =\frac V\hbar \int d\tau \left[ i\frac{M_0}%
\gamma \left( \frac{d\phi }{d\tau }\right) -i\frac{M_0}\gamma \left( \frac{%
d\phi }{d\tau }\right) \cos \theta +E\left( \theta ,\phi \right) \right] , 
\eqnum{2}
\end{equation}
where $V$ is the volume of the FM particle and $\gamma $ is the gyromagnetic
ratio. $M_0=\left| \overrightarrow{M}\right| =\hbar \gamma S/V$, where $S$
is the total spin of FM particles. It is noted that the first two terms in
Eq. (2) define the topological Berry or Wess-Zumino, Chern-Simons term which
arises from the nonorthogonality of spin coherent states. The Wess-Zumino
term has a simple topological interpretation. For a closed path, this term
equals $-iS$ times the area swept out on the unit sphere between the path
and the north pole. The first term in Eq. (2) is a total imaginary-time
derivative, which has no effect on the classical equations of motion, but it
is crucial for the spin-parity effects.\cite{1,11,12,13,14,15} However, for
the closed instanton or bounce trajectory described in this paper (as shown
in the following), this time derivative gives a zero contribution to the
path integral, and therefore can be omitted.

In the semiclassical limit, the dominant contribution to the transition
amplitude comes from finite action solution (instanton or bounce) of the
classical equation of motion. The instanton's contribution to the tunneling
rate $\Gamma $ or the tunnel splitting $\Delta $ (not including the
topological Wess-Zumino phase) is given by\cite{1} 
\begin{equation}
\Gamma \ (\text{or }\ \Delta )=A\omega _p\left( \frac{S_{cl}}{2\pi }\right)
^{1/2}e^{-S_{cl}},  \eqnum{7}
\end{equation}
where $\omega _p$ is the oscillation frequency in the well, $S_{cl}$ is the
classical action, and the prefactor $A$ originates from the quantum
fluctuations about the classical path. It is noted that Eq. (7) is based on
quantum tunneling at the level of ground state, and the temperature
dependence of the tunneling frequency is not taken into account. However,
the instanton technique is suitable only for the evaluation of the tunneling
rate or the tunnel splitting at the vacuum level, since the usual (vacuum)
instantons satisfy the vacuum boundary conditions. Recently, Liang {\it et al%
}.\cite{9,10} developed new types of pseudoparticle configurations which
satisfy periodic boundary condition (i.e., periodic instantons or nonvacuum
instantons). They found that the tunneling effect indeed increases
exponentially with energy in the low-energy region.

For a particle moving in a double-well-like potential $U\left( x\right) $,
the level splittings of degenerate excited levels or the imaginary parts of
the metastable levels at an energy $E>0$ are given by the following formula
in the WKB approximation,\cite{6,16,17} 
\begin{equation}
\Delta E\text{ }\left( \text{or }%
%TCIMACRO{\func{Im}}
%BeginExpansion
\mathop{\rm Im}%
%EndExpansion
E\right) =\frac{\omega \left( E\right) }\pi \exp \left[ -S\left( E\right)
\right] ,  \eqnum{8}
\end{equation}
and the imaginary-time action is 
\begin{equation}
S\left( E\right) =2\sqrt{2m}\int_{x_1\left( E\right) }^{x_2\left( E\right)
}dx\sqrt{U\left( x\right) -E},  \eqnum{9}
\end{equation}
where $x_{1,2}\left( E\right) $ are the turning points for the particle
oscillating inside the inverted potential $-U\left( x\right) $. $\omega
\left( E\right) =2\pi /t\left( E\right) $ is the energy-dependent frequency,
and $t\left( E\right) $ is the period of the real-time oscillation in the
potential well, 
\begin{equation}
t\left( E\right) =\sqrt{2m}\int_{x_3\left( E\right) }^{x_4\left( E\right) }%
\frac{dx}{\sqrt{E-U\left( x\right) }},  \eqnum{10}
\end{equation}
where $x_{3,4}\left( E\right) $ are the turning points for the particle
oscillating inside the potential $U\left( x\right) $. Recently, the
crossover from quantum to classical behavior and the associated phase
transition were studied extensively in nanospin systems\cite{6,17,18,19,20}
and other systems.\cite{21}

\section*{III. MQC and MQT for biaxial crystal symmetry}

In this section, we study the tunneling behaviors of the magnetization
vector in single-domain FM nanoparticle which has the biaxial crystal
symmetry, with $\pm \widehat{z}$ being the easy axes in the absence of an
external magnetic field. The magnetic field is applied in the $ZX$ plane, at
an angle in the range of $\pi /2\leq \theta _H\leq \pi $. Then the
magnetocrystalline anisotropy energy $E\left( \theta ,\phi \right) $ can be
written as 
\begin{equation}
E\left( \theta ,\phi \right) =K_1\sin ^2\theta +K_2\sin ^2\theta \sin ^2\phi
-M_0H_x\sin \theta \cos \phi -M_0H_z\cos \theta +E_0,  \eqnum{11}
\end{equation}
where $K_1$ and $K_2$ are the longitudinal and the transverse anisotropy
coefficients, respectively, and $E_0$ is a constant which makes $E\left(
\theta ,\phi \right) $ zero at the initial orientation. As the external
magnetic field is applied in the $ZX$ plane, $H_x=H\sin \theta _H$ and $%
H_z=H\cos \theta _H$, where $H$ is the magnitude of the field and $\theta _H$
is the angle between the magnetic field and the $\widehat{z}$ axis.

By introducing the dimensionless parameters as 
\begin{equation}
\overline{K}_2=K_2/2K_1,\overline{H}_x=H_x/H_0,\overline{H}_z=H_z/H_0, 
\eqnum{12}
\end{equation}
the $E\left( \theta ,\phi \right) $ term of Eq. (11) can be rewritten as 
\begin{equation}
\overline{E}\left( \theta ,\phi \right) =\frac 12\sin ^2\theta +\overline{K}%
_2\sin ^2\theta \sin ^2\phi -\overline{H}_x\sin \theta \cos \phi -\overline{H%
}_z\cos \theta +\overline{E}_0,  \eqnum{13}
\end{equation}
where $E\left( \theta ,\phi \right) =2K_1\overline{E}\left( \theta ,\phi
\right) $, and $H_0=2K_1/M_0$. At finite magnetic field, the plane given by $%
\phi =0$ is the easy plane, on which $\overline{E}\left( \theta ,\phi
\right) $ reduces to 
\begin{equation}
\overline{E}\left( \theta ,\phi =0\right) =\frac 12\sin ^2\theta -\overline{H%
}\cos \left( \theta -\theta _H\right) +\overline{E}_0.  \eqnum{14}
\end{equation}

We denote $\theta _0$ to be the initial angle and $\theta _c$ the critical
angle at which the energy barrier vanishes when the external magnetic field
is close to the critical value $\overline{H}_c\left( \theta _H\right) $ (to
be calculated in the following). Then, $\theta _0$ satisfies $\left[ d%
\overline{E}\left( \theta ,\phi =0\right) /d\theta \right] _{\theta =\theta
_0}=0$, $\theta _c$ and $\overline{H}_c$ satisfy both $\left[ d\overline{E}%
\left( \theta ,\phi =0\right) /d\theta \right] _{\theta =\theta _c,\overline{%
H}=\overline{H}_c}=0$ and $\left[ d^2\overline{E}\left( \theta ,\phi
=0\right) /d\theta ^2\right] _{\theta =\theta _c,\overline{H}=\overline{H}%
_c}=0$, which leads to 
\begin{eqnarray}
\frac 12\sin \left( 2\theta _0\right) +\overline{H}\sin \left( \theta
_0-\theta _H\right) &=&0,  \eqnum{15a} \\
\frac 12\sin \left( 2\theta _c\right) +\overline{H}_c\sin \left( \theta
_c-\theta _H\right) &=&0,  \eqnum{15b} \\
\cos \left( 2\theta _c\right) +\overline{H}_c\cos \left( \theta _c-\theta
_H\right) &=&0.  \eqnum{15c}
\end{eqnarray}
After some algebra, the dimensionless critical field $\overline{H}_c\left(
\theta _H\right) $ and the critical angle $\theta _c$ are found to be 
\begin{eqnarray}
\overline{H}_c &=&\left[ \left( \sin \theta _H\right) ^{2/3}+\left| \cos
\theta _H\right| ^{2/3}\right] ^{-3/2},  \eqnum{16a} \\
\sin \left( 2\theta _c\right) &=&\frac{2\left| \cot \theta _H\right| ^{1/3}}{%
1+\left| \cot \theta _H\right| ^{2/3}}.  \eqnum{16b}
\end{eqnarray}

Now we consider the limiting case that the external magnetic field is
slightly lower than the critical field, i.e., $\epsilon =1-\overline{H}/%
\overline{H}_c\ll 1$. At this practically interesting situation, the barrier
height is low and the width is narrow, and therefore the tunneling rate in
MQT or the tunnel splitting in MQC is large. Introducing $\eta \equiv \theta
_c-\theta _0$ $\left( \left| \eta \right| \ll 1\text{ in the limit of }%
\epsilon \ll 1\right) $, expanding $\left[ d\overline{E}\left( \theta ,\phi
=0\right) /d\theta \right] _{\theta =\theta _0}=0$ about $\theta _c$, and
using the relations $\left[ d\overline{E}\left( \theta ,\phi =0\right)
/d\theta \right] _{\theta =\theta _c,\overline{H}=\overline{H}_c}=0$ and $%
\left[ d^2\overline{E}\left( \theta ,\phi =0\right) /d\theta ^2\right]
_{\theta =\theta _c,\overline{H}=\overline{H}_c}=0$, Eq. (15a) becomes 
\begin{equation}
\sin \left( 2\theta _c\right) \left( \epsilon -\frac 32\eta ^2\right) -\eta
\cos \left( 2\theta _c\right) \left( 2\epsilon -\eta ^2\right) =0. 
\eqnum{17}
\end{equation}
Then the potential energy $\overline{E}\left( \theta ,\phi \right) $ reduces
to the following equation in the limit of small $\epsilon $, 
\begin{equation}
\overline{E}\left( \delta ,\phi \right) =\overline{K}_2\sin ^2\phi \sin
^2\left( \theta _0+\delta \right) +\overline{H}_x\sin \left( \theta
_0+\delta \right) \left( 1-\cos \phi \right) +\overline{E}_1\left( \delta
\right) ,  \eqnum{18}
\end{equation}
where $\delta \equiv \theta -\theta _0$ $\left( \left| \delta \right| \ll 1%
\text{ in the limit of }\epsilon \ll 1\right) $, and $\overline{E}_1\left(
\delta \right) $ is a function of only $\delta $ given by 
\begin{equation}
\overline{E}_1\left( \delta \right) =\frac 14\sin \left( 2\theta _c\right)
\left( 3\delta ^2\eta -\delta ^3\right) +\frac 12\cos \left( 2\theta
_c\right) \left[ \delta ^2\left( \epsilon -\frac 32\eta ^2\right) +\delta
^3\eta -\frac 14\delta ^4\right] .  \eqnum{19}
\end{equation}

In the following, we will investigate the tunneling behaviors of the
magnetization vector at excited levels in single-domain FM nanoparticles
with biaxial crystal symmetry at three different angle ranges of the
external magnetic field as $\theta _H=\pi /2$, $\pi /2+O\left( \epsilon
^{3/2}\right) <\theta _H<\pi -O\left( \epsilon ^{3/2}\right) $, and $\theta
_H=\pi $, respectively.

\subsection*{A. $\theta _H=\pi /2$}

For $\theta _H=\pi /2$, we have $\theta _c=\pi /2$ from Eq. (16b) and $\eta =%
\sqrt{2\epsilon }$ from Eq. (17). Eqs (18) and (19) show that $\phi $ is
very small for the full range of angles $\pi /2\leq \theta _H\leq \pi $ for
FM particles with biaxial crystal symmetry. Performing the Gaussian
integration over $\phi $, we can map the spin system onto a particle moving
problem in the one-dimensional potential well. Now the imaginary-time
transition amplitude Eqs. (1) and (2) becomes 
\begin{eqnarray}
U_{fi} &=&\int d\delta \exp \left( -S_E\left[ \delta \right] \right) , 
\nonumber \\
&=&\int d\delta \exp \left\{ -\int d\tau \left[ \frac 12m\left( \frac{%
d\delta }{d\tau }\right) ^2+U\left( \delta \right) \right] \right\} , 
\eqnum{20}
\end{eqnarray}
with 
\[
m=\frac{\hbar S^2}{2V\left[ K_2+K_1\left( 1-\epsilon \right) \right] }, 
\]
and 
\begin{equation}
U\left( \delta \right) =\frac{K_1V}{4\hbar }\delta ^2\left( \delta -2\sqrt{%
2\epsilon }\right) ^2.  \eqnum{21}
\end{equation}
The problem is one of MQC, where the magnetization vector resonates
coherently between the energetically degenerate easy directions at $\delta
=0 $ and $\delta =2\sqrt{2\epsilon }$ separated by a classically
impenetrable barrier at $\delta =\sqrt{2\epsilon }$.

Now we apply the periodic instanton method\cite{9,10} to evaluate the level
splittings of excited states. The periodic (or thermal) instanton
configuration $\delta _p$ which minimizes the Euclidean action in Eq. (20)
satisfies the equation of motion 
\begin{equation}
\frac 12m\left( \frac{d\delta _p}{d\tau }\right) ^2-U\left( \delta _p\right)
=-E,  \eqnum{22}
\end{equation}
where $E>0$ is a constant of integration, which can be viewed as the
classical energy of the pseudoparticle configuration. Then the kink-solution
is\cite{10} 
\begin{equation}
\delta _p=\sqrt{2\epsilon }+\sqrt{2\epsilon -\alpha }\text{sn}\left( \omega
_1\tau ,k\right) ,  \eqnum{23}
\end{equation}
where $\alpha =2\sqrt{\hbar E/K_1V}$, and $\omega _1=\sqrt{K_1V/2\hbar m}%
\sqrt{2\epsilon +\alpha }$. sn$\left( \omega _1\tau ,k\right) $ is the
Jacobian elliptic sine function of modulus $k=\sqrt{\left( 2\epsilon -\alpha
\right) /\left( 2\epsilon +\alpha \right) }$. In the low energy limit, i.e., 
$E\rightarrow 0$, $k\rightarrow 1$, sn$\left( u,1\right) \rightarrow \tanh u$%
, we have 
\begin{equation}
\delta _p=\sqrt{2\epsilon }\left[ 1+\tanh \left( \widetilde{\omega }_1\tau
\right) \right] ,  \eqnum{24}
\end{equation}
which is exactly the vacuum instanton solution derived in Ref. 4, where $%
\widetilde{\omega }_1=\sqrt{2}\left( K_1V/\hbar S\right) \epsilon ^{1/2}%
\sqrt{1-\epsilon +K_2/K_1}$.

The Euclidean action of the periodic instanton configuration Eq. (23) over
the domain $\left( -\beta ,\beta \right) $ is found to be 
\begin{equation}
S_p=\int_{-\beta }^\beta d\tau \left[ \frac 12m\left( \frac{d\delta _p}{%
d\tau }\right) ^2+U\left( \delta _p\right) \right] =W+2E\beta ,  \eqnum{25}
\end{equation}
with 
\begin{equation}
W=\frac 1{3\sqrt{2}}\sqrt{\frac{K_1Vm}\hbar }\left( 2\epsilon \right) ^{3/2}%
\frac 1{\sqrt{1-k^{\prime 2}/2}}\left[ E\left( k\right) -\frac{k^{\prime 2}}{%
2-k^{\prime 2}}K\left( k\right) \right] ,  \eqnum{26}
\end{equation}
where $k^{\prime 2}=1-k^2$. $K\left( k\right) $ and $E\left( k\right) $ are
the complete elliptic integral of the first and second kind, respectively.
The general formula Eq. (8) gives the tunnel splittings of excited levels as 
\begin{equation}
\Delta E=\frac{\omega \left( E\right) }\pi \exp \left( -W\right) , 
\eqnum{27}
\end{equation}
where $W$ is shown in Eq. (26), and $\omega \left( E\right) =2\pi /t\left(
E\right) $ is the energy-dependent frequency. For this case, the period $%
t\left( E\right) $ is found to be 
\begin{equation}
t\left( E\right) =\sqrt{2m}\int_{\delta _1}^{\delta _2}\frac{d\delta }{\sqrt{%
E-U\left( \delta \right) }}=\sqrt{\frac{2\hbar m}{K_1V}}\frac 1{\sqrt{%
2\epsilon +\alpha }}K\left( k^{\prime }\right) ,  \eqnum{28}
\end{equation}
where $\delta _1=\sqrt{2\epsilon }+\sqrt{2\epsilon -\alpha }$, and $\delta
_2=\sqrt{2\epsilon }+\sqrt{2\epsilon +\alpha }$. Now we discuss the low
energy limit where $E$ is much less than the barrier height. In this case, $%
k^{\prime 4}=4\hbar E/K_1V\epsilon ^2\ll 1$, so we can perform the
expansions of $K\left( k\right) $ and $E\left( k\right) $ in Eq. (26) to
include terms like $k^{\prime 4}$ and $k^{\prime 4}\ln \left( 4/k^{\prime
}\right) $, 
\begin{eqnarray}
E\left( k\right)  &=&1+\frac 12\left[ \ln \left( \frac 4{k^{\prime }}\right)
-\frac 12\right] k^{\prime 2}+\frac 3{16}\left[ \ln \left( \frac 4{k^{\prime
}}\right) -\frac{13}{12}\right] k^{\prime 4}\cdots ,  \nonumber \\
K\left( k\right)  &=&\ln \left( \frac 4{k^{\prime }}\right) +\frac 14\left[
\ln \left( \frac 4{k^{\prime }}\right) -1\right] k^{\prime 2}+\frac 9{64}%
\left[ \ln \left( \frac 4{k^{\prime }}\right) -\frac 76\right] k^{\prime
4}\cdots .  \eqnum{29}
\end{eqnarray}
With the help of small oscillator approximation for energy near the bottom
of the potential well, $E_n=\left( n+1/2\right) \Omega _1$, $\Omega _1=\sqrt{%
U^{\prime \prime }\left( \delta =\sqrt{2\epsilon }\right) /m}=2\sqrt{%
K_1V\epsilon /\hbar m}$, Eq. (26) is expanded as 
\begin{equation}
W=\frac 83\sqrt{\frac{K_1Vm}\hbar }\epsilon ^{3/2}-\left( n+\frac 12\right)
+\left( n+\frac 12\right) \ln \left[ \frac 1{64\epsilon ^{3/2}}\sqrt{\frac %
\hbar {K_1Vm}}\left( n+\frac 12\right) \right] .  \eqnum{30}
\end{equation}
Then the general formula Eq. (8) gives the low-lying energy shift of $n$-th
excited states for FM particles with biaxial crystal symmetry in the
presence of an external magnetic field applied perpendicular to the
anisotropy axis $\left( \theta _H=\pi /2\right) $ as 
\begin{eqnarray}
\hbar \Delta E_n &=&\frac 2{n!\sqrt{\pi }}\left( K_1V\right) \epsilon
^{1/2}S^{-1}\sqrt{1-\epsilon +\lambda }\left( \frac{2^{11/2}\epsilon ^{3/2}S%
}{\sqrt{1-\epsilon +\lambda }}\right) ^{n+1/2}  \nonumber \\
&&\times \exp \left( -\frac{2^{5/2}}3\frac S{\sqrt{1-\epsilon +\lambda }}%
\epsilon ^{3/2}\right) ,  \eqnum{31}
\end{eqnarray}
where $\lambda =K_2/K_1$.

When $n=0$, the energy shift of the ground state is 
\begin{equation}
\hbar \Delta E_0=\frac{2^{15/4}}{\sqrt{\pi }}\left( K_1V\right) \left(
1-\epsilon +\lambda \right) ^{1/4}\epsilon ^{5/4}S^{-1/2}\exp \left( -\frac{%
2^{5/2}}3\frac S{\sqrt{1-\epsilon +\lambda }}\epsilon ^{3/2}\right) . 
\eqnum{32}
\end{equation}
Then Eq. (31) can be written as 
\begin{equation}
\hbar \Delta E_n=\frac{q_1^n}{n!}\left( \hbar \Delta E_0\right) ,  \eqnum{33}
\end{equation}
where 
\begin{equation}
q_1=\frac{2^{11/2}\epsilon ^{3/2}S}{\sqrt{1-\epsilon +\lambda }}.  \eqnum{34}
\end{equation}
To see the temperature dependence we take the Boltzmann average of the
tunneling frequency $f=4\Delta E$ at temperature $T$, 
\begin{equation}
f\left( T\right) =\frac 1{Z_0}\sum_n4\Delta E_n\exp \left( -\hbar E_n\beta
\right) ,  \eqnum{35}
\end{equation}
where $Z_0=\sum_n\exp \left( -\hbar E_n\beta \right) $ is the partition
function with the harmonic oscillator approximated eigenvalues $E_n=\left(
n+1/2\right) \Omega _1$. The final result of the tunneling frequency at a
finite temperature $T$ is found to be 
\begin{equation}
f\left( T\right) =4\Delta E_0\left( 1-e^{-\hbar \Omega _1\beta }\right) \exp
\left( q_1e^{-\hbar \Omega _1\beta }\right) ,  \eqnum{36}
\end{equation}
where $\Delta E_0$ and $q_1$ are shown in Eqs. (32) abd (34).

\subsection*{B. $\pi /2+O\left( \epsilon ^{3/2}\right) <\theta _H<\pi
-O\left( \epsilon ^{3/2}\right) $}

For $\pi /2+O\left( \epsilon ^{3/2}\right) <\theta _H<\pi -O\left( \epsilon
^{3/2}\right) $, the critical angle $\theta _c$ is in the range of $O\left(
\epsilon ^{3/2}\right) <\theta _c<\pi /2-O\left( \epsilon ^{3/2}\right) $,
and $\eta \approx \sqrt{2\epsilon /3}$. Now the problem can be mapped onto a
problem of one-dimensional motion by integrating out $\phi $, and for this
case the effective mass $m$ and the potential $U\left( \delta \right) $ in
Eq. (20) are found to be 
\[
m=\frac{\hbar S^2}{2K_1V\left[ \frac{1-\epsilon }{1+\left| \cot \theta
_H\right| ^{2/3}}+\lambda \right] }, 
\]
and 
\begin{eqnarray}
U\left( \delta \right) &=&\frac{K_1V}{2\hbar }\sin 2\theta _c\left( \sqrt{%
6\epsilon }\delta ^2-\delta ^3\right)  \nonumber \\
&=&3U_0q^2\left( 1-\frac 23q\right) ,  \eqnum{37}
\end{eqnarray}
where $q=3\delta /2\sqrt{6\epsilon }$, and $U_0=\left(
2^{5/2}/3^{3/2}\right) \left( K_1V\epsilon ^{3/2}/\hbar \right) \sin 2\theta
_c$. The problem becomes one of MQT, where the magnetization vector escapes
from the metastable state at $\delta =0$, $\phi =0$ through the barrier by
quantum tunneling.

Now the periodic bounce configuration with an energy $E>0$ is found to be 
\begin{equation}
\delta _p=\frac 23\sqrt{6\epsilon }\left[ a-\left( a-b\right) \text{sn}%
^2\left( \omega _2\tau ,k\right) \right] ,  \eqnum{38}
\end{equation}
where 
\begin{equation}
\omega _2=\frac 1{2^{1/4}\times 3^{1/4}}\sqrt{\frac{K_1V}{\hbar m}}\sqrt{%
\sin 2\theta _c}\epsilon ^{1/4}\sqrt{a-c}.  \eqnum{39}
\end{equation}
$a\left( E\right) >b\left( E\right) >c\left( E\right) $ denote three roots
of the cubic equation 
\begin{equation}
q^3-\frac 32q^2+\frac E{2U_0}=0.  \eqnum{40}
\end{equation}
sn$\left( \omega _2\tau ,k\right) $ is the Jacobian elliptic sine function
of modulus $k=\sqrt{\left( a-b\right) /\left( a-c\right) }$. In the low
energy limit, i.e., $E\rightarrow 0$, $k\rightarrow 1$, sn$\left( u,1\right)
\rightarrow \tanh u$, $a\rightarrow 3/2$, $b\rightarrow 0$, we have 
\begin{equation}
\delta _p=\frac{\sqrt{6\epsilon }}{\cosh ^2\left( \widetilde{\omega }_2\tau
\right) },  \eqnum{41}
\end{equation}
where 
\[
\widetilde{\omega }_2=\frac{3^{1/4}}{2^{1/4}}\left( \frac{K_1V}{\hbar S}%
\right) \epsilon ^{1/4}\frac{\left| \cot \theta _H\right| ^{1/6}}{1+\left|
\cot \theta _H\right| ^{2/3}}\sqrt{1-\epsilon +\lambda \left( 1+\left| \cot
\theta _H\right| ^{2/3}\right) }.
\]
Eq. (41) agrees well with the vacuum bounce solution obtained in Ref. 4.

The classical action of the periodic bounce configuration Eq. (38) is 
\begin{equation}
S_p=\int_{-\beta }^\beta d\tau \left[ \frac 12m\left( \frac{d\delta _p}{%
d\tau }\right) ^2+U\left( \delta _p\right) \right] =W+2E\beta ,  \eqnum{42}
\end{equation}
with 
\begin{equation}
W=\frac{2^{9/2}}{5\times 3^{3/2}}\sqrt{m\epsilon U_0}\left( a-c\right)
^{5/2}\left[ 2\left( k^4-k^2+1\right) E\left( k\right) -\left( 1-k^2\right)
\left( 2-k^2\right) K\left( k\right) \right] .  \eqnum{43}
\end{equation}
The period $t\left( E\right) $ of this case is found to be 
\begin{equation}
t\left( E\right) =\sqrt{2m}\int_c^b\frac{d\delta }{\sqrt{E-U\left( \delta
\right) }}=4\sqrt{\frac{2\epsilon m}{3U_0\left( a-c\right) }}K\left(
k^{\prime }\right) ,  \eqnum{44}
\end{equation}
where $k^{\prime 2}=1-k^2$. Then the general formula Eq. (8) gives the
imaginary parts of the metastable energy levels as 
\begin{equation}
%TCIMACRO{\func{Im}}
%BeginExpansion
\mathop{\rm Im}%
%EndExpansion
E=\frac{\omega \left( E\right) }\pi \exp \left( -W\right) ,  \eqnum{45}
\end{equation}
where $\omega \left( E\right) =2\pi /t\left( E\right) $, and $W$ is shown in
Eq. (43).

Here we discuss the low energy limit of the imaginary part of the metastable
energy levels. For this case, $E_n=\left( n+1/2\right) \Omega _2$, $\Omega
_2=\sqrt{U^{\prime \prime }\left( \delta =0\right) /m}=\left( 3/2\right) 
\sqrt{U_0/m\epsilon }$, $a\approx \left( 3/2\right) \left( 1-k^{\prime
2}/4\right) $, $b\approx \left( 3k^{\prime 2}/4\right) \left( 1+3k^{\prime
2}/4\right) $, $c\approx -\left( 3k^{\prime 2}/4\right) \left( 1+k^{\prime
2}/4\right) $, and $k^{\prime 4}=16E/27U_0\ll 1$. Therefore, Eqs (43) and
(45) reduce to 
\begin{eqnarray}
W &=&\frac{2^{17/4}\times 3^{1/4}}5S\epsilon ^{5/4}\frac{\left| \cot \theta
_H\right| ^{1/6}}{\sqrt{1-\epsilon +\lambda \left( 1+\left| \cot \theta
_H\right| ^{2/3}\right) }}-\left( n+\frac 12\right)   \nonumber \\
&&+\left( n+\frac 12\right) \ln \left[ \frac{2^{25/4}\times
3^{11/4}S\epsilon ^{5/4}\left| \cot \theta _H\right| ^{1/6}}{\left( n+\frac 1%
2\right) \sqrt{1-\epsilon +\lambda \left( 1+\left| \cot \theta _H\right|
^{2/3}\right) }}\right] ,  \eqnum{46}
\end{eqnarray}
and 
\begin{eqnarray}
\hbar 
%TCIMACRO{\func{Im}}
%BeginExpansion
\mathop{\rm Im}%
%EndExpansion
E_n &=&\frac{3^{1/4}\times 2^{3/4}}{n!\sqrt{\pi }}\epsilon
^{1/4}S^{-1}\left( K_1V\right) \frac{\left| \cot \theta _H\right| ^{1/6}}{%
1+\left| \cot \theta _H\right| ^{2/3}}\sqrt{1-\epsilon +\lambda \left(
1+\left| \cot \theta _H\right| ^{2/3}\right) }  \nonumber \\
&&\times \left( \frac{2^{25/4}\times 3^{11/4}S\epsilon ^{5/4}\left| \cot
\theta _H\right| ^{1/6}}{\sqrt{1-\epsilon +\lambda \left( 1+\left| \cot
\theta _H\right| ^{2/3}\right) }}\right) ^{n+1/2}  \nonumber \\
&&\times \exp \left( -\frac{2^{17/4}\times 3^{1/4}}5S\epsilon ^{5/4}\frac{%
\left| \cot \theta _H\right| ^{1/6}}{\sqrt{1-\epsilon +\lambda \left(
1+\left| \cot \theta _H\right| ^{2/3}\right) }}\right) .  \eqnum{47}
\end{eqnarray}
For vacuum bounce case $n=0$, we have 
\begin{eqnarray}
\hbar 
%TCIMACRO{\func{Im}}
%BeginExpansion
\mathop{\rm Im}%
%EndExpansion
E_0 &=&\frac{3^{13/9}\times 2^{31/8}}{\sqrt{\pi }}\left( K_1V\right)
\epsilon ^{7/8}S^{-1/2}\frac{\left| \cot \theta _H\right| ^{1/6}}{1+\left|
\cot \theta _H\right| ^{2/3}}\left[ 1-\epsilon +\lambda \left( 1+\left| \cot
\theta _H\right| ^{2/3}\right) \right] ^{1/4}  \nonumber \\
&&\times \exp \left( -\frac{2^{17/4}\times 3^{1/4}}5S\epsilon ^{5/4}\frac{%
\left| \cot \theta _H\right| ^{1/6}}{\sqrt{1-\epsilon +\lambda \left(
1+\left| \cot \theta _H\right| ^{2/3}\right) }}\right) .  \eqnum{48}
\end{eqnarray}
Then Eq. (47) can be written as 
\begin{equation}
\hbar 
%TCIMACRO{\func{Im}}
%BeginExpansion
\mathop{\rm Im}%
%EndExpansion
E_n=\frac{q_2^n}{n!}\left( \hbar 
%TCIMACRO{\func{Im}}
%BeginExpansion
\mathop{\rm Im}%
%EndExpansion
E_0\right) ,  \eqnum{49}
\end{equation}
where 
\begin{equation}
q_2=\frac{2^{25/4}\times 3^{11/4}S\epsilon ^{5/4}\left| \cot \theta
_H\right| ^{1/6}}{\sqrt{1-\epsilon +\lambda \left( 1+\left| \cot \theta
_H\right| ^{2/3}\right) }}.  \eqnum{50}
\end{equation}
At finite temperature $T$ the decay rate $\Gamma =2%
%TCIMACRO{\func{Im}}
%BeginExpansion
\mathop{\rm Im}%
%EndExpansion
E$ can be easily found by averaging over the Boltzmann distribution 
\begin{equation}
\Gamma \left( T\right) =\frac 2{Z_0}\sum_n%
%TCIMACRO{\func{Im}}
%BeginExpansion
\mathop{\rm Im}%
%EndExpansion
E_n\exp \left( -E_n\beta \right) ,  \eqnum{51}
\end{equation}
where $Z_0=\sum_n\exp \left( -\hbar E_n\beta \right) $ is the partition
function with the harmonic oscillator approximated eigenvalues $E_n=\left(
n+1/2\right) \Omega _2$. The final result of the decay rate at a finite
temperature $T$ is found to be 
\begin{equation}
\Gamma \left( T\right) =2%
%TCIMACRO{\func{Im}}
%BeginExpansion
\mathop{\rm Im}%
%EndExpansion
E_0\left( 1-e^{-\hbar \Omega _2\beta }\right) \exp \left( q_2e^{-\hbar
\Omega _2\beta }\right) ,  \eqnum{52}
\end{equation}
where $%
%TCIMACRO{\func{Im}}
%BeginExpansion
\mathop{\rm Im}%
%EndExpansion
E_0$ and $q_2$ are shown in Eqs. (48) abd (50).

In Fig. 1 we plot the temperature dependence of the tunneling rate for the
typical values of parameters for nanometer-scale single-domain ferromagnets: 
$S=5000$, $\epsilon =1-\overline{H}/\overline{H}_c=2\times 10^{-3},$ $%
\lambda =K_2/K_1=10$, and $\theta _H=3\pi /4$. From Fig. 1 we easily see the
crossover from purely quantum tunneling to thermally assisted quantum
tunneling. The temperature $T_0^{\left( 0\right) }$ characterizing the
crossover from quantum to thermal regimes can be estimated as $%
k_BT_0^{\left( 0\right) }=\Delta U/S_0$, where $\Delta U$ is the barrier
height, and $S_0$ is the WKB\ exponent of the ground-state tunneling. It is
shown that in the cubic potential $\left( q^2-q^3\right) $, the usual
second-order phase transition from the thermal to the quantum regimes occurs
as the temperature is lowered.\cite{17} The second-order phase transition
temperature is given by $k_BT_0^{\left( 2\right) }=\hbar \omega _b/2\pi $,
where $\omega _b=\sqrt{\left| U^{\prime \prime }\left( x_b\right) \right| /m}
$ is the frequency of small oscillations near the bottom of the inverted
potential $-U\left( x\right) $, and $x_b$ corresponds to the bottom of the
inverted potential. For the present case, $\delta _b=2\sqrt{6\epsilon }/3$, 
\[
\hbar \omega _b=2^{5/4}\times 3^{1/4}\left( K_1V\right) S^{-1}\epsilon ^{1/4}%
\frac{\left| \cot \theta _H\right| ^{1/6}}{1+\left| \cot \theta _H\right|
^{2/3}}\sqrt{1-\epsilon +\lambda \left( 1+\left| \cot \theta _H\right|
^{2/3}\right) },
\]
\[
S_0=\frac{2^{17/4}\times 3^{1/4}}5S\epsilon ^{5/4}\frac{\left| \cot \theta
_H\right| ^{1/6}}{\sqrt{1-\epsilon +\lambda \left( 1+\left| \cot \theta
_H\right| ^{2/3}\right) }},
\]
and 
\[
\hbar \Delta U=\frac{2^{7/2}}{3^{3/2}}\left( K_1V\right) \epsilon ^{3/2}%
\frac{\left| \cot \theta _H\right| ^{1/6}}{1+\left| \cot \theta _H\right|
^{2/3}}.
\]
Then it is easy to obtain that 
\[
k_BT_0^{\left( 2\right) }=\frac{2^{1/4}\times 3^{1/4}}\pi \left( K_1V\right)
S^{-1}\epsilon ^{1/4}\frac{\left| \cot \theta _H\right| ^{1/6}}{1+\left|
\cot \theta _H\right| ^{2/3}}\sqrt{1-\epsilon +\lambda \left( 1+\left| \cot
\theta _H\right| ^{2/3}\right) },
\]
and $k_BT_0^{\left( 0\right) }=\left( 5\pi /18\right) k_BT_0^{\left(
2\right) }\approx 0.87k_BT_0^{\left( 2\right) }$.

\subsection*{C. $\theta _H=\pi $}

In case of $\theta _H=\pi $, we have $\theta _c=0$ and $\eta =0$. Working
out the integration over $\phi $, the spin tunneling problem is mapped onto
the problem of a particle with effective mass $m=\hbar S^2/2K_2V$ moving in
the one dimensional potential well $U\left( \delta \right) =\left(
K_1V/\hbar \right) \left( \epsilon \delta ^2-\delta ^4/4\right) $. Now the
problem is one of MQT, and the nonvacuum bounce at a given energy $E>0$ is
found to be 
\begin{equation}
\delta _p=\sqrt{2\epsilon }\left( 1+\sqrt{1-\frac{\hbar E}{K_1V\epsilon ^2}}%
\right) ^{1/2}\text{dn}\left( \omega _3\tau ,k\right) ,  \eqnum{53}
\end{equation}
where 
\begin{eqnarray*}
\omega _3 &=&\sqrt{\frac{K_1V}{\hbar m}}\left( 1+\sqrt{1-\frac{\hbar E}{%
K_1V\epsilon ^2}}\right) ^{1/2}, \\
k^2 &=&1-\left( \frac{1-\sqrt{1-\frac{\hbar E}{K_1V\epsilon ^2}}}{1+\sqrt{1-%
\frac{\hbar E}{K_1V\epsilon ^2}}}\right) ^2.
\end{eqnarray*}
In the loe energy limit, i.e., $E\rightarrow 0$, $k\rightarrow 1$, dn$\left(
u,1\right) \rightarrow 1/\cosh u$, we have 
\begin{equation}
\delta _p=\frac{2\sqrt{\epsilon }}{\cosh \left( \widetilde{\omega }_3\tau
\right) },  \eqnum{54}
\end{equation}
where $\widetilde{\omega }_3=2\sqrt{K_1K_2\epsilon }V/\hbar S$. Eq. (54) is
in good agreement with the vacuum bounce solution derived in Ref. 4.

The classical action of the nonvacuum bounce Eq. (53) is 
\begin{equation}
S_p=\int_{-\beta }^\beta d\tau \left[ \frac 12m\left( \frac{d\delta _p}{%
d\tau }\right) ^2+U\left( \delta _p\right) \right] =W+2E\beta ,  \eqnum{55}
\end{equation}
with 
\begin{equation}
W=\frac 43m\epsilon \left( 1+\sqrt{1-\frac{\hbar E}{K_1V\epsilon ^2}}\right)
^2\omega _3\left[ \left( 2-k^2\right) E\left( k\right) -2k^{\prime 2}K\left(
k\right) \right] ,  \eqnum{56}
\end{equation}
where $k^{\prime 2}=1-k^2$. Then the imaginary parts of the metastable
energy levels are 
\begin{equation}
%TCIMACRO{\func{Im}}
%BeginExpansion
\mathop{\rm Im}%
%EndExpansion
E=\frac{\omega \left( E\right) }\pi \exp \left( -W\right) ,  \eqnum{57}
\end{equation}
where $\omega \left( E\right) =2\pi /t\left( E\right) $, and the period $%
t\left( E\right) $ for this case is found to be 
\begin{equation}
t\left( E\right) =4\sqrt{\frac{\hbar m}{K_1V\epsilon }}\frac 1{1+\sqrt{1-%
\frac{\hbar E}{K_1V\epsilon ^2}}}K\left( k^{\prime }\right) .  \eqnum{58}
\end{equation}

Now we consider the low energy limit of the imaginary part of the metastable
energy level. For this case, $E_n=\left( n+1/2\right) \Omega _3$, $\Omega _3=%
\sqrt{U^{\prime \prime }\left( \delta =0\right) /m}=\sqrt{2K_1V\epsilon
/\hbar m}$, $k^{\prime 2}=\left( 1/2^{3/2}\epsilon ^{3/2}\right) \sqrt{\hbar
/K_1Vm}\left( n+1/2\right) \ll 1$, then 
\begin{equation}
W=\frac 83\sqrt{\frac{K_1}{K_2}}S\epsilon ^{3/2}-\left( n+\frac 12\right)
-\left( n+\frac 12\right) \ln \left[ \frac{32S\epsilon ^{3/2}\sqrt{K_1/K_2}}{%
n+1/2}\right] ,  \eqnum{59}
\end{equation}
and 
\begin{equation}
\hbar 
%TCIMACRO{\func{Im}}
%BeginExpansion
\mathop{\rm Im}%
%EndExpansion
E_n=\frac{\sqrt{2}}{n!\sqrt{\pi }}\left( K_1V\right) \lambda
^{1/2}S^{-1}\epsilon ^{1/2}\left( \frac{32\epsilon ^{3/2}S}{\sqrt{\lambda }}%
\right) ^{n+1/2}\exp \left( -\frac 83\frac{S\epsilon ^{3/2}}{\sqrt{\lambda }}%
\right) .  \eqnum{60}
\end{equation}
In the case of $n=0$, the imaginary part of the metastable ground state
reduces to 
\begin{equation}
\hbar 
%TCIMACRO{\func{Im}}
%BeginExpansion
\mathop{\rm Im}%
%EndExpansion
E_0=\frac 8{\sqrt{\pi }}\left( K_1V\right) S^{-1/2}\epsilon ^{5/4}\exp
\left( -\frac 83\frac{S\epsilon ^{3/2}}{\sqrt{\lambda }}\right) .  \eqnum{61}
\end{equation}
Then Eq. (60) can be written as 
\begin{equation}
\hbar 
%TCIMACRO{\func{Im}}
%BeginExpansion
\mathop{\rm Im}%
%EndExpansion
E_n=\frac{q_3^n}{n!}\left( \hbar 
%TCIMACRO{\func{Im}}
%BeginExpansion
\mathop{\rm Im}%
%EndExpansion
E_0\right) ,  \eqnum{62}
\end{equation}
where 
\begin{equation}
q_3=\frac{32\epsilon ^{3/2}S}{\sqrt{\lambda }}.  \eqnum{63}
\end{equation}
And the final result of the decay rate at finite temperature $T$ is found to
be 
\begin{equation}
\Gamma \left( T\right) =2%
%TCIMACRO{\func{Im}}
%BeginExpansion
\mathop{\rm Im}%
%EndExpansion
E_0\left( 1-e^{-\hbar \Omega _3\beta }\right) \exp \left( q_3e^{-\hbar
\Omega _3\beta }\right) .  \eqnum{64}
\end{equation}

The temperature dependence of the decay rate is shown in Fig. 2. It can be
shown that the double-well potential $\left( q^2-q^4\right) $ yields the
second-order phase transition from the thermal to the quantum regimes as the
temperature is lowered. For this case, the position of the energy barrier is 
$\delta _b=\sqrt{2\epsilon }$, the frequency of small oscillations near the
bottom of the inverted potential is $\hbar \omega _b=2^{3/2}\left( \sqrt{%
K_1K_2}V\right) \epsilon ^{1/2}S^{-1}$, the WKB exponent of the ground-state
tunneling is $S_0=\left( 8/3\sqrt{\lambda }\right) S\epsilon ^{3/2}$, and
the height of barrier is $\hbar \Delta U=\left( K_1V\right) \epsilon ^2$.
Therefore, $k_BT_0^{\left( 2\right) }=\left( \sqrt{2}/\pi \right) \left( 
\sqrt{K_1K_2}V\right) \epsilon ^{1/2}S^{-1}$, and $k_BT_0^{\left( 0\right)
}=\left( 3\pi /8\sqrt{2}\right) k_BT_0^{\left( 2\right) }\approx
0.83k_BT_0^{\left( 2\right) }$.

\section*{IV. Conclusions}

In summary we have investigated the quantum tunneling of the magnetization
vector between excited levels in single-domain FM nanoparticles with biaxial
crystal symmetry in the presence of an external magnetic field at
arbitrarily directed angle. By applying the periodic instanton method in the
spin-coherent-state path-integral representation, we obtain the analytic
formulas for the tunnel splitting between degenerate excited levels in MQC
and the imaginary parts of the metastable energy levels in MQT of the
magnetization vector in the low barrier limit for the external magnetic
field perpendicular to the easy axis $\left( \theta _H=\pi /2\right) $, for
the field antiparallel to the initial easy axis $\left( \theta _H=\pi
\right) $, and for the field at an angle between these two orientations $%
\left( \pi /2+O\left( \epsilon ^{3/2}\right) <\theta _H<\pi -O\left(
\epsilon ^{3/2}\right) \right) $. The temperature dependences of the
tunneling frequency and the decay rate are clearly shown for each case. One
important conclusion is that the tunneling rate and the tunnel splitting at
excited levels depend on the orientation of the external magnetic field
distinctly. Even a small misalignment of the field with $\theta _H=\pi /2$
and $\pi $ orientations can completely change the results of the tunneling
rates. Another interesting conclusion concerns the field strength dependence
of the WKB\ exponent in the tunnel splitting or the tunneling rate. It is
found that in a wide range of angles, the $\epsilon \left( =1-\overline{H}/%
\overline{H}_c\right) $ dependence of the WKB exponent is given by $\epsilon
^{5/4}$, not $\epsilon ^{3/2}$ for $\theta _H=\pi /2$, and $\theta _H=\pi $.
As a result, we conclude that both the orientation and the strength of the
external magnetic field are the controllable parameters for the experimental
test of the phenomena of macroscopic quantum tunneling and coherence of the
magnetization vector between excited levels in single-domain FM
nanoparticles at sufficiently low temperatures. The theoretical calculations
performed in this paper can be extended to the FM\ particles with a much
more complex structure of magnetocrystalline anisotropy energy, such as
trigonal, tetragonal, and hexagonal crystal symmetries. Work along this line
is still in progress. We hope that the theoretical results presented in this
paper may stimulate more experiments whose aim is observing macroscopic
quantum tunneling and coherence in nanometer-scale single-domain
ferromagnets.

\section*{Acknowledgments}

R.L. would like to acknowledge Dr. Hui Hu, Dr. Yi Zhou, Dr. Jian-She Liu,
Professor Zhan Xu, Professor Jiu-Qing Liang and Professor Fu-Cho Pu for
stimulating discussions. R. L. and J. L. Zhu would like to thank Professor
W. Wernsdorfer and Professor R. Sessoli for providing their paper (Ref. 15).

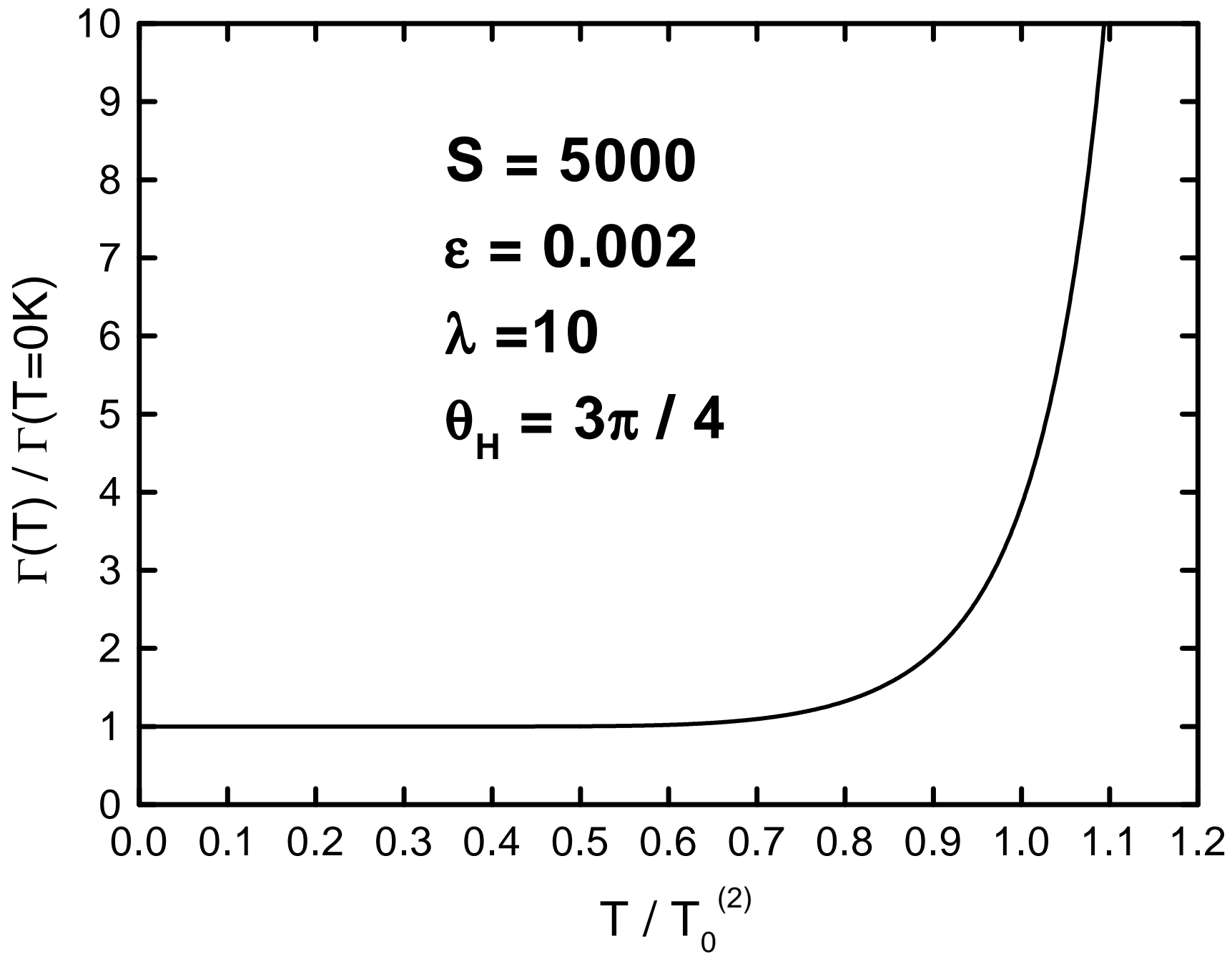
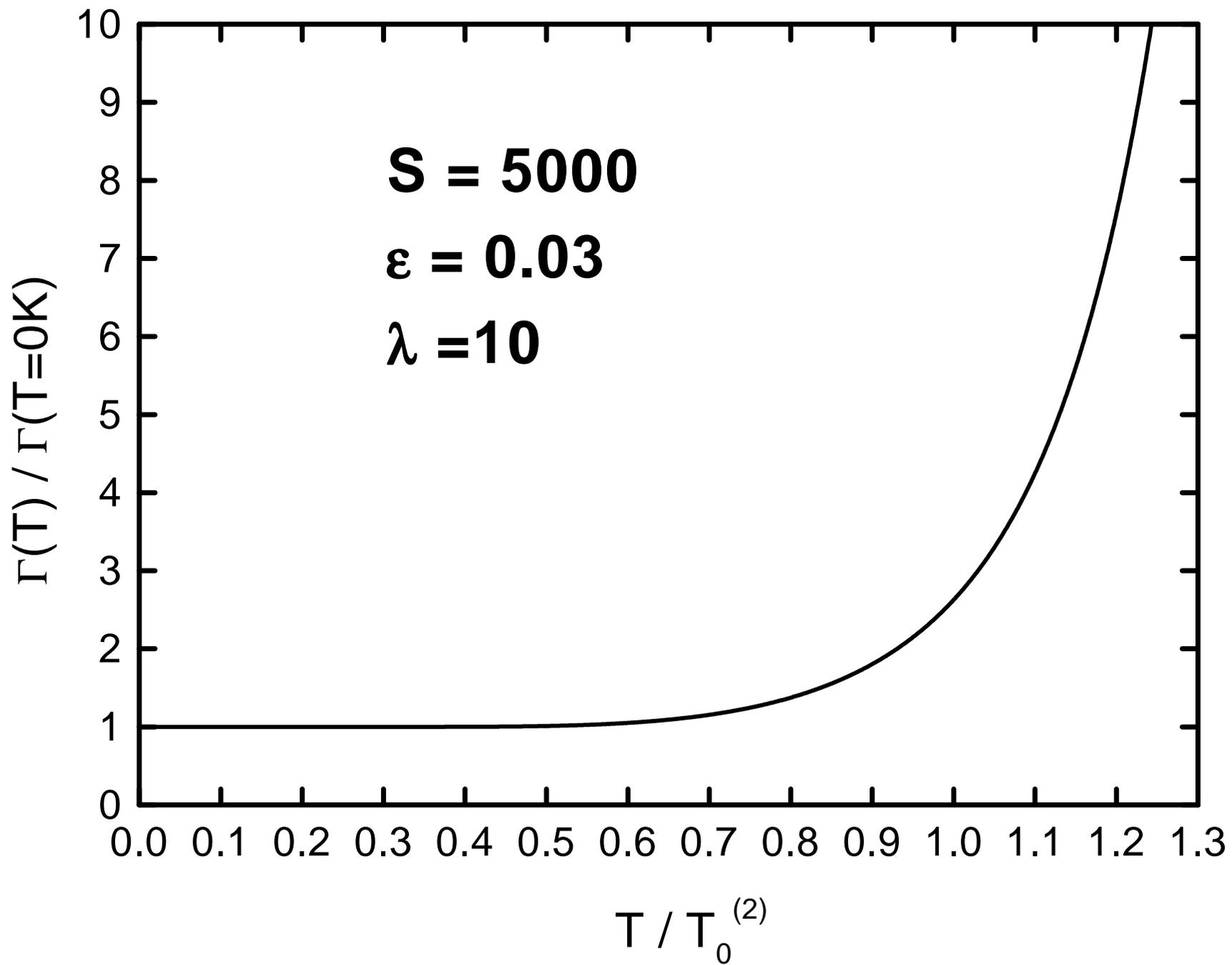
\end{document}